\documentclass[prd,twocolumn,showpacs,showkeys]{revtex4} 

 \usepackage{amsmath}
 \usepackage{graphicx}
 \usepackage{amssymb}
 \usepackage{amsfonts}
 \usepackage{latexsym}

\def\beq{\begin{eqnarray}}
\def\eeq{\end{eqnarray}}
\def\ln{\,\mbox{ln}\,}
\def\Ln{\,\mbox{Ln}\,}
\def\Det{\,\mbox{Det}\,}
\def\det{\,\mbox{det}\,}
\def\tr{\,\mbox{tr}\,}
\def\diag{\,\mbox{diag}\,}
\def\Tr{\,\mbox{Tr}\,}
\def\sTr{\,\mbox{sTr}\,}

\def\al{\alpha}
\def\be{\beta}

\def\ga{\gamma}
\def\de{\delta}

\def\ep{\epsilon}
\def\ze{\zeta}

\def\la{\lambda}
\def\na{\nabla}
\def\pa{\partial}
\def\ro{\varrho}

\def\si{\sigma}
\def\om{\omega}
\def\ph{\varphi}
\def\ta{\tau}
\def\th{\theta}

\def\Ga{\Gamma}
\def\De{\Delta}

\def\EA{\,effective action\,}

\begin{document}

\title{One-loop corrections to the 
photon propagator in the curved-space QED}

\author{Bruno Gon\c{c}alves}
\email{brunoxgoncalves@yahoo.com.br}
\affiliation{Departamento de F\'{\i}sica, ICE,
Universidade Federal de Juiz de Fora,
Juiz de Fora, CEP: 36036-330, MG,  Brazil}

\author{Guilherme de Berredo-Peixoto}
\email{guilherme@fisica.ufjf.br}
\affiliation{Departamento de F\'{\i}sica, ICE,
Universidade Federal de Juiz de Fora,
Juiz de Fora, CEP: 36036-330, MG,  Brazil}

\author{Ilya L. Shapiro}
\email{shapiro@fisica.ufjf.br}
\affiliation{Departamento de F\'{\i}sica, ICE,
Universidade Federal de Juiz de Fora,
Juiz de Fora, CEP: 36036-330, MG,  Brazil }
\altaffiliation[Also at ]{Tomsk State Pedagogical 
University, Tomsk, Russia.}

\begin{abstract}

We calculate and discuss the one-loop 
corrections to the photon sector of QED interacting to a 
background gravitational field. At high energies the fermion 
field can be taken as massless and the quantum terms can be 
obtained by integrating conformal anomaly. We present a 
covariant local expression for the corresponding effective 
action, similar to the one obtained earlier for the gravitational 
sector. At the moderate energies the quantum terms can be obtained
through the heat-kernel method. In this way we derive the exact
one-loop $\be$-function for the electric charge in the momentum 
subtraction scheme and explore both massless and large-mass 
limits. The relation between the two approaches is shown and 
the difference discussed in view of the possible applications
to cosmology and astrophysics. 
\end{abstract}
\pacs{04.62.+v; 
11.15.Kc;   
11.10.Hi; 	
12.20.-m    
}
\keywords{ Photon propagation,
QED, formfactors, beta functions, conformal anomaly}

\maketitle

\section{Introduction}

The interaction of electromagnetic field with gravity is 
an important subject due to various astrophysical and 
cosmological applications. There are many publications on 
the subject and, in particular, many interesting works devoted 
to the quantum corrections in the electromagnetic sector of 
the theory. In the early publications \cite{Milton,Gastmans}, 
direct calculations were performed through the Feynman diagrams 
and Schwinger source method and later on, starting from 
\cite{Drummond}, by using different modifications of the 
heat-kernel approach 
\cite{Mazzitelli,Shore,Avramidi_QED-2009,Bastianelli}.
The diagrams were also used in the traditional style 
calculations \cite{Latorre,Mann} in flat space-time and 
using a modified approach admitting a generalization 
to the curved space-time \cite{Prokopec}\footnote{One more 
approach to break conformal symmetry is based on the 
spontaneous breaking of Lorentz invariance \cite{Mota}.}. 
A general consideration, including one and two-loop 
calculations and taking into account the temperature 
effects, has been given in \cite{DG}. 

One of the most important features of quantum corrections 
to the action of electromagnetic field is that they break 
the conformal invariance which the theory possesses at the 
classical level. Let us start by presenting a short list 
of the cosmological and astrophysical situations where the 
violation of conformal symmetry can be relevant. 
\
(1) The breaking of conformal symmetry changes the equation 
of state of the radiation, that can be relevant in the 
radiation-dominated epoch in the early universe. The modified 
equation of state for the radiation enables one to construct 
interesting cosmological models (see, e.g., \cite{bounce} and
references therein), including the ones \cite{Winfried} based 
on the quantum corrections of \cite{Drummond}. 
The modified equation of state can produce the change in the 
law of expansion of the universe, entropy production and 
other phenomena. Also, this effect can slightly affect the 
red-shift dependence of both energy density of radiation 
and cosmic microwave background (CMB) temperature. Because 
of the growing precision of 
astrophysical experiments, at some point this feature of 
quantum corrections can be relevant.  
\
(2) As a particular aspect of the previous point, due to the 
broken conformal symmetry the rate of creation
of the photons in the reheating period after inflation may be 
affected by quantum corrections, leading to the potentially 
observable consequences. 
\
(3) The violation of conformal symmetry is needed for the 
creation of initial seeds of magnetic field of the galaxies, 
at the epoch of structure formation \cite{WT}. There is a 
well-known attempt to explain this violation by quantum effects 
similar to conformal anomaly \cite{Dolgov93}, during the 
inflationary epoch (see also \cite{Mazzi}). It would 
be rather interesting to have a more complete 
understanding of the field theoretical 
mechanisms behind such violation, including the ones which can 
occur at much later epochs. The problem was further discussed, 
e.g., in \cite{magnet,Calzetta}, and one can also see 
\cite{Rubinstein-review} and \cite{Giovannini-review} 
for the recent reviews of the possible origin of the initial 
seeds of cosmic magnetic fields and related subjects. 
\ 
(4) Furthermore, the quantum corrections to the photon 
propagator can modify the position of the pole and hence 
produce the situation when the electromagnetic wave 
propagates with the velocity which is slightly different 
from the one in the purely classical case. A similar effect 
can take place in curved space and is sometimes 
characterized as a superluminal motion 
\cite{Superluminal,Superluminal2}
(see \cite{Shore-review} for the recent review). In 
particular, this effect may have a significant impact 
on the behavior of light in the vicinity of the black 
hole \cite{bh1,bh2,bh3}. Indeed, a similar effect may 
also take place due to the presence of the boundaries 
and, in general, because of the macroscopic conditions 
\cite{Macro}. Let us remark that the light propagation
can be affected also by the nonlinear terms \cite{nonlinear} 
which can result from the quantum corrections and by the 
classical and/or quantum interactions to external gravity 
\cite{Obukhov} or to other fields such as $k$-essence
\cite{Vikman}.
It is obvious that the conformal symmetry forbids most of
the possible changes in the wave equations and therefore 
the detailed study of the conformal symmetry breaking is
relevant for this issue too. 

If thinking about conformal symmetry, the first remarkable
observation is that in the realistic theories such as QED 
or the standard model (SM), and also in its generalizations 
such as supersymmetric standard model or grand unified theories, 
the massless 
and conformal invariant electromagnetic field couples to the 
other fields, which are all massive and therefore conformally 
noninvariant. In the present paper we shall concentrate on 
the simplest case of QED and try to present the most general 
view of the violation of conformal symmetry within this theory. 
However, there is no qualitative difference other mentioned 
theories, which can be also considered elsewhere. 

One can formulate the two main questions: 
\\
1) What is the 
mechanism of violation of the local conformal symmetry
in the intermediate energies and how one can link its 
violations at the ultraviolet (UV) and infrared (IR) limit. 
Is it possible to 
derive such relations? This is not a purely technical 
question, because during its evolution the Universe 
is passing different phases and it is desirable to 
know how the conformal symmetry is violated not only 
at the inflation epoch or at the present epoch, but 
also around the recombination epoch or at the time 
period when the cosmic structure starts to form. The 
main purpose of the present paper is to explore this 
issue. 
\\
2) To which extent the finite 
quantum corrections and, in particular, the violation  
of local conformal symmetry, are universal? In other 
words, do we have some ambiguity in the quantum terms?
Answering the last question is of course very significant,
because the effective action, in general, is not a 
uniquely defined object. Usually, it depends on the 
choice of parametrization of quantum fields, in particular 
on the gauge fixing choice, on the calculational schemes, 
regularization, renormalization, etc. 

Our purpose is to derive the most general expression for 
the one-loop quantum correction to the electromagnetic 
sector of QED in curved space-time. As we have already 
mentioned above, the physical situations for such quantum 
corrections in the early universe and at the later period
are very different. In the first case the fermions can 
be treated as almost massless. In this case the mechanism 
for the violation of conformal symmetry is the well-known 
conformal anomaly. The advantage of conformal anomaly as
a method of deriving quantum corrections to the classical 
effects is its simplicity, direct relation to the UV 
divergencies and consequent universality. At the opposite 
end of the energy scale the masses of the virtual fermions 
are much greater than the energies of the real photons or, 
equivalently, the energies of the external tails in the 
loops of massive fermions. In this case we observe the 
decoupling of the quantum contributions of the massive 
fields, according to the Appelquist and Carazzone theorem 
\cite{AC}. The violation of conformal symmetry still exists, 
but it is related to the remnant higher derivative terms 
in the effective action, which are quadratically 
suppressed by the fermion masses. Also, in curved 
space-time, there may be curvature-dependent terms which 
can break down the local conformal symmetry. Since we 
intend to study the curved-space theory, we will always 
concentrate our attention on the local conformal symmetry. 
 
The paper is organized as follows. In the next section we 
consider the simplified version of QED with a massless 
fermion and obtain the noncovariant and also nonlocal 
covariant forms of the anomaly-induced effective action. 
Some part of this section was previously known, but we 
present these results for the sake of completeness, and 
also to establish the most general local representation for 
the covariant expression. In Sect. 3 we start to consider 
massive case and calculate the first three 
coefficients of the Schwinger-DeWitt expansion in an 
arbitrary space-time dimension $D$. Even though the focus 
of our interest is mainly on the $4D$ case, it proves 
useful to have a more general $D$-dimensional 
result. The calculations are performed in two 
different schemes, such that we can try the limits
of universality of the quantum corrections. The scheme
dependence of the one-loop effective action is explored 
in a parallel paper \cite{multi}. In Sect. 4 we use the 
heat-kernel method to calculate the complete one-loop 
formfactors for the electromagnetic field sector. In the 
same section we consider the UV limit and establish the 
relation of the massless limit for the one-loop corrections 
for the massive case with the ones derived via conformal 
anomaly. In Sect. 5 we consider the renormalization 
group equations for the massive case and derive different 
forms of the low-energy decoupling law corresponding to 
the distinct calculational schemes. In Sect. 6 we analyze 
the corresponding running of the effective charge and compare
it to the one in the minimal-subtraction renormalization 
scheme. Finally, in the last section we present some 
discussions and draw our conclusions.

\section{Anomaly-induced action for the metric 
and electromagnetic fields background}

Let us start from a brief survey of massless conformal 
QED. The theory should be formulated in curved space-time 
and hence the action depends on the electromagnetic 
potential $A_\mu$, Dirac spinor field $\psi$ and on the 
external metric $g_{\mu\nu}$. As far as we are interested 
in the local conformal symmetry, one of the useful 
parametrizations of the metric is 
\beq
{g}_{\mu\nu} = {\bar g}_{\mu\nu}\cdot e^{2\si}\,,
\qquad \si=\si(x)\,,
\label{conf}
\eeq
where $\,{\bar g}_{\mu\nu}\,$ is the fiducial metric with 
fixed determinant. For example, in the case of the
cosmological metric, using spherical coordinates, we have 
\beq
{\bar g}_{\mu\nu}\,=\,\diag\,
\Big(1,\,-\,\frac{1}{1-kr^2},\,\,-r^2\sin^2\th,\,\,-r^2
\Big)\,.
\nonumber
\eeq

Separating $\si(x)$ in (\ref{conf}) proves to be a useful
tool, especially because of the relation 
\beq
\frac{2}{\sqrt{g}}\,g_{\mu\nu}
\frac{\de\,A[g_{\mu\nu}]}{\de\, g_{\mu\nu}}
= \frac{e^{- 4\si}}{\sqrt{{\bar g}}}
\left.\frac{\de\,A[{\bar g}_{\mu\nu}\,e^{2\si}]}{\de \si}
\,\right|_{{\bar g_{\mu\nu}}\rightarrow g_{\mu\nu},
\si\rightarrow 0}\,,
\label{deriv}
\eeq  
which is valid for any functional $\,A[g_{\mu\nu}]\,$ 
of the metric and maybe other fields. 
If we replace the action of some theory in curved space 
at the place of $\,A[g_{\mu\nu}]$, then the {\it l.h.s.} 
of the above relation is nothing else but the trace of 
the corresponding energy-momentum tensor, $T^\mu_\mu$. 
In order to remove the effect of other field variables, 
it is sufficient to use the corresponding equations of 
motion.  

The vanishing trace of the energy-momentum tensor implies 
that the conformal factor of the metric decouples from the 
matter. However, the situation changes dramatically if we 
take quantum effects onto account. In the last case the 
corresponding theoretical phenomenon is called the trace 
anomaly \cite{duff77} (see also \cite{duff94,PoS-Conform} 
for review and many further references).  

The classical action of electromagnetic field is
\beq
S_{em}\,=\,-\frac14\,\int d^4x 
\sqrt{g}\,\, F_{\mu\nu} \, F^{\mu\nu}
\label{em}
\eeq
and possesses local conformal invariance. The last 
means this action does not change under simultaneous 
transformation of the metric and of the vector $A_\mu$,
namely
\beq
g_{\mu\nu} \to g_{\mu\nu}^\prime = g_{\mu\nu}\,e^{2\si}
\,,\quad
A_\mu \to A_\mu^\prime = A_\mu\,.
\label{vec trans}
\eeq
Let us note that the difference between conformal 
weight and dimension for the vector field is due to
the vector field definition in curved space-time,
\beq
A_\mu = A_b\,e^b_\mu
\,,\quad
e^b_\mu\,e^a_\nu\,\eta_{ab}=g_{\mu\nu}\,,
\quad
e^b_\mu\,e^a_\nu\,g^{\mu\nu}=\eta^{ab}\,.
\label{vec-only}
\eeq

We are interested in the corrections to the action (\ref{em}) 
due to quantum effects of the fermion 
\beq
S_{f}\,=\,i\,\int d^4x\sqrt{g}\,\left\{\,
\bar{\psi}\,\ga^\mu\,\big(\na_{\mu} - ieA_{\mu}\big)\psi 
\,-\, im\,\bar{\psi}\,\psi\,\right\}\,.
\label{spi}
\eeq
The conformal transformation rule for spinors is
\beq
\psi \to \psi^\prime=\psi\,e^{-3\si/2}
\,,\quad
{\bar \psi} \to {\bar \psi}^\prime={\bar \psi}\,e^{-3\si/2}\,.
\nonumber
\eeq
The metric is always transformed like in (\ref{vec trans}).
Indeed the action (\ref{spi}) is conformal invariant only 
when the spinor mass is zero, $m=0$. All those fermions 
which couple to the electromagnetic field 
as in (\ref{spi}) are massive, however, the relevance of 
the mass terms depends on the energy scale. For instance, 
if we are interested in the quantum effects of fermions 
close to the inflationary epoch, the kinetic energy of the 
real fermions and (more important) of photons is much
greater than the mass of the fermions. In this situation 
taking a massless spinor field is a legitimate approximation, 
so let us start from this case and take $m=0$. 

The derivation of conformal anomaly in the presence of 
background metric and electromagnetic field has been 
discussed before \cite{duff77,duff94}, and we can use 
the known result. The conformal anomaly can be used to 
construct the equation for the finite part of the one-loop 
correction to the effective action of the background 
metric and electromagnetic potential,
\beq
\,T^\mu_\mu \,&=&\,
\frac{2}{\sqrt{g}}\,g_{\mu\nu}
\frac{\de\, {\bar \Ga}_{ind}}{\de g_{\mu\nu}}
\,
\cr
&=&\, \frac{1}{(4\pi)^2}\,
\left(\,wC^2 + bE + c{\Box} R + \tilde{\be} F_{\mu\nu}^2 
\,\right)\,,
\label{mainequation}
\eeq
where 
$$
C^2=C_{\mu\nu\al\be}^2=R_{\mu\nu\al\be}^2 - 2 R_{\al\be}^2 + (1/3)\,R^2
$$
is the square of the Weyl tensor and 
$$
E = R_{\mu\nu\al\be}^2 - 4 R_{\al\be}^2 + R^2
$$
is the integrand of the Gauss-Bonnet topological term.
The coefficients $\,\om,\,b,\,c\,$ depend on the
number of scalar $N_s$, fermion $N_f$ and massless 
vector $N_v$ fields as
\beq
\om  &=& \frac{1}{120}\,N_s + \frac{1}{20}\,N_f + 
\frac{1}{10}\,N_v\,,
\nonumber
\\
b &=& -\,\frac{1}{360}\,N_s 
- \frac{11}{360}\,N_f - \frac{31}{180}\,N_v \,,
\nonumber
\\
c &=& \frac{1}{180}\,N_s
+ \frac{1}{30}\,N_f - \frac{1}{10}\,N_v \,.
\label{abc}
\eeq
Also, $\tilde{\be}$ depends on the number of charged
scalars (in case of scalar QED) and spinors. As far as
the massless approximation can be applied in the very 
early universe, the number of fields which contribute 
to these coefficients is not necessarily restricted by 
the QED framework. 

The solution of Eq. (\ref{mainequation})
is straightforward \cite{rei}
(see also generalizations for the theory with torsion 
\cite{buodsh} and with a scalar field \cite{shocom}). The 
simplest possibility is to parametrize metric as in
(\ref{conf}), separating the conformal factor 
\ $\sigma(x)$ \ and rewrite Eq. (\ref{mainequation}) 
using (\ref{deriv}). The solution for the effective 
action is
\beq
{\bar \Ga} &=& S_c[{\bar g}_{\mu\nu},\,A_\mu]
\,\,+\,\, \frac{1}{(4\pi)^2}\,
\int d^4 x\sqrt{{\bar g}}\,\Big\{ 
\om\si {\bar C}^2 
\cr
&+& \tilde{\be} \si {\bar F}_{\mu\nu}^2 
+ b\si ({\bar E}-\frac23 {\bar {\Box}}
{\bar R})
+  2b\si{\bar \De}_4\si 
\cr
&-& \frac{1}{12}\,(c+\frac23 b)[{\bar R} - 6({\bar \na}\si)^2 
- ({\bar \Box} \si)]^2\Big\}
\label{quantum}
\eeq
where $S_c[{\bar g}_{\mu\nu},\,A_\mu]=S_c[g_{\mu\nu}]$ 
is an unknown conformal invariant functional of the metric 
and $\,A_\mu$, which serves as an integration constant for 
the Eq. (\ref{mainequation}). All quantities with bars are
constructed using the metric ${\bar g}_{\mu\nu}$, in 
particular 
$$
{\bar F}_{\mu\nu}^2={\bar F}_{\mu\nu}{\bar F}_{\al\be}
{\bar g}^{\mu\al}{\bar g}^{\be\nu}\,.
$$
Furthermore, $\De_4$ is the fourth derivative conformally 
covariant operator acting on dimensionless scalar 
\beq
\De_4 \,=\, \Box^2 + 2R^{\mu\nu}\na_\mu\na_\nu
- \frac23\,R\Box +\frac13\,R_{;\mu}\,\na^\mu\,.
\label{rei op}
\eeq

The solution (\ref{quantum}) has the merit of being simple, 
but an important disadvantage is that it is not covariant or, 
in other words, it is not expressed in terms of an original 
metric $\,g_{\mu\nu}$. In order to obtain the nonlocal 
covariant solution and after represent it in the local 
form using auxiliary fields, we shall follow \cite{rei,a}. 
The presence of the ${\bar F}^2_{\mu\nu}$ terms does not 
require any essential changes compared to the consideration
presented in \cite{PoS-Conform}, in particular this term 
can be always taken together with the ${\bar C}^2$ one. 
So, we present just the final result in the nonlocal 
form, which is expressed in terms of the Green function 
$G(x,y)$ of the operator (\ref{rei op}),
$$
\De_{4,x}\,G(x,y)=\de(x,y)\,.
$$

Using the last formulas and (\ref{deriv}) we find, for any 
\ $A(g_{\mu\nu}) = A\big({\bar g}_{\mu\nu}\,e^{2\si}\big)$, 
\ the relation
\beq
\frac{\de}{\de \si (y)}\,\int &d^4 x& \sqrt{g (x)}\,A\,
\left.
\Big(E - \frac23{\Box}R\Big)
\right|_{g_{\mu\nu} \,=\, {\bar g}_{\mu\nu}}\,=
\cr
\cr
&=&\,
4\sqrt{{\bar g}}{\bar {\De}}_4 \,A 
\,=\, 4\sqrt{g}{\De}_4 \,A \,.
\label{GB}
\eeq
In particular, we obtain
\beq 
\Gamma_{induced}=\Ga_\om + \Ga_b + \Ga_c\,,
\label{nonloc}
\eeq
where 
\begin{widetext}
\beq
\Gamma_\om \, = \,\frac14\,\int d^4 x \sqrt{g (x)}\, 
\int d^4 y \sqrt{-g (y)}\, 
\Big(\om C^2 + \tilde{\be} F_{\mu\nu}^2 \Big)_x
\,G(x,y)\,\Big(E - \frac23{\Box}R \Big)_y\,,
\label{a-term}
\eeq
\beq
\Gamma_b = \frac{b}{8}\,\int d^4x \sqrt{g (x)}\, 
\int d^4 y \sqrt{g (y)}\,
\Big(E - \frac23{\Box}R \Big)_x\,G(x,y)\,\Big(E - \frac23{\Box}R \Big)_y
\label{b-term}
\eeq
and 
\beq
\Ga_{c} = - \frac{c+\frac23\,b}{12(4\pi)^2}
\,\int d^4 x \sqrt{g (x)}\,R^2(x) \,.
\label{c-term}
\eeq
\end{widetext}

The nonlocal expressions for the anomaly-induced 
effective action can be presented in a local form 
using two auxiliary scalar fields $\ph$ and $\psi$
\cite{a}. 
Let us give just a final result which has an extra 
electromagnetic terms compared to the one described 
in \cite{PoS-Conform}
\beq
\Gamma &=& S_c[g_{\mu\nu},\,A_\la] 
- \frac{3c+2b}{36(4\pi)^2}\,\int d^4 x \sqrt{g (x)}\,R^2(x) 
\cr
&+&\,  \int d^4 x \sqrt{g (x)}\,\Big\{
\frac12 \,\ph\De_4\ph - \frac12 \,\psi\De_ 4\psi
\nonumber
\\
&+& \ph\,\left[\,\frac{\sqrt{-b}}{8\pi}\,(E -\frac23\,{\Box}R)\,
- \frac{1}{8\pi\sqrt{-b}}\,
\left(aC^2+\tilde{\be} F_{\mu\nu}^2\right)\,\right] 
\cr
&+& 
\frac{1}{8\pi \sqrt{-b}}\,\psi\,\left(aC^2+\tilde{\be} 
F_{\mu\nu}^2\right) \,\Big\}\,.
\label{finaction}
\eeq
The local covariant form (\ref{finaction}) is dynamically 
equivalent to the nonlocal covariant form (\ref{nonloc}). 
The complete 
definition of the Cauchy problem in the theory with the 
nonlocal action requires defining the boundary conditions for 
the Green functions \ $G(x,y)$, which show up independently 
in the two terms \ (\ref{a-term}) \ and \ (\ref{b-term}). The 
same can be achieved, in the local version, by imposing the 
boundary conditions on the two auxiliary fields $\ph$ 
and $\psi$. 

Let us separate the part of the effective action 
(\ref{finaction}), which has direct relation to the 
electromagnetic terms and therefore represents a one-loop
correction for the classical action (\ref{em}),
\beq
\Gamma &=& S_c[g_{\mu\nu},A_\la] 
\,\,+\,\, \int d^4 x \sqrt{g (x)}\,\,\Big\{
\frac{\sqrt{-b}}{8\pi}\,\ph\,\Big(E -\frac23{\Box}R\Big)
\label{em-1-loop}
\cr
&+&  \frac12  \int d^4 x \sqrt{g (x)}\Big[
\ph\De_4\ph - \psi\De_ 4\psi
\cr
&+& \frac{1}{4\pi \sqrt{-b}}\,(\psi-\ph)\,
\tilde{\be} F_{\mu\nu}^2 \Big].
\eeq
Let us note that the presence of the $A_\mu$-independent 
term $\ph\big[E -(2/3){\Box}R\big]$ is relevant, because 
only this term provides violation of local conformal 
symmetry in the whole expression. For instance, the 
connection with the noncovariant presentation 
(\ref{quantum}) is through the relations
\beq
\sqrt{{\bar g}}\,{\bar \De}_4 &=& \sqrt{g}\,{\De}_4
\qquad
\mbox{and}
\cr\cr
\sqrt{g}\Big(E - \frac23{\Box}R \Big) &=& \sqrt{{\bar g}}
\Big({\bar E}
- \frac23{\bar {\Box}}{\bar R} + 4{\bar {\De}}_4\si \Big)\,.
\label{trans Gauss}
\eeq
Another important observation is that the anomalous 
metric dependence of the $F_{\mu\nu}^2$ in Eq. 
(\ref{em-1-loop}) appears due to the coupling of 
$F_{\mu\nu}^2$ with the auxiliary scalars $\psi$ and
$\ph$. This dependence is nontrivial because these 
two fields have different space-time behavior due 
to their distinct dynamical equations and independent 
initial and boundary conditions. 

As an example of the use of the auxiliary fields $\psi$ and 
$\ph$, we mention that the different choices of their boundary 
conditions enable one to classify the vacuum states for the 
semiclassical black hole \cite{balsan}. 
It would be interesting to explore the initial and 
boundary conditions for $\psi$ and $\ph$ such that it 
could be applied, for example, for calculating the creation 
of the seeds of a magnetic field during inflation. However 
this issue requires (and deserves) a special detailed 
investigation which goes beyond the scope of the present 
paper. 


\section{Massive fermion case: Schwinger-DeWitt expansion}

The results of the previous section can be seen as follows: 
in the case of massless and conformal fields we can calculate 
an important (sometimes the most important) part of the 
one-loop \EA by using conformal anomaly. The qualitative 
explanation of this possibility is that the conformal 
anomaly is essentially controlled by the logarithmic 
divergences of the theory. And, on the other hand, 
the logarithmic divergences are intimately related to the 
UV behavior of the theory, or, better to say, to its reaction 
to the scaling in the high energy region. 
In the massless theory there is no natural scale and, 
therefore, all kind of quantum processes can be actually 
seen as ``high energy''. For this reason the UV divergences 
provide much more information on the finite part of \EA in 
the massless conformal case than they do for massive fields. 
The question is what we can do in this case. 

The theory of a massive quantum field 
has the natural scale established by the
mass. In this situation to notions of ``high energy region''
(UV) and ``low energy region'' (IR) assume some relations 
between the energy of the field and 
the mass of the quantum field. The quantum effects of 
massive fields are supposed to be close to the ones of 
the massless fields in the UV and follow the decoupling 
theorem \cite{AC} in the IR. In both cases the most 
relevant finite part of \EA is given by nonlocal 
expressions, but the structure of nonlocalities are 
rather different in the two cases. As far as we are 
interested in the photon propagation, all that we need 
is a formfactor of the electromagnetic term. We shall 
present this result, at the one-loop level, in the next 
section. In the present section we consider the local 
terms which correspond to the Schwinger-DeWitt expansion 
of the \EA in the massive case. The analysis of the 
coefficients of this expansion will prove useful for 
better understanding of the structure of the formfactors 
of our interest, moreover deriving these coefficients      
provides an efficient check of correctness of the 
consequent general calculations. 

The one-loop effective action (EA) in the metric and 
electromagnetic sectors can be defined via the path 
integral 
\beq
e^{i \Gamma[g_{\mu\nu},\,A_\mu]}=
\int{D\psi D\bar{\psi}\, e^{i S_{QED}}}\,,
\label{eaction}
\eeq
where 
\beq
S_{QED}\,=\,S_{f}+S_{em}
\label{total}
\eeq
and the actions $S_{f}$, $S_{em}$ are defined in (\ref{em})
and (\ref{spi}). Since the action $S_{QED}$ is bilinear in 
the spinor fields, we find (see, e.g., \cite{book})
\beq
{\bar \Ga}^{(1)} &=& \,-\,\frac{1}{2}\,\Ln \Det\,\hat{H}\,,
\label{TrLog}
\eeq
where
\beq
\hat{H} &=& i
\big(\ga^\mu\na_\mu - im - ie \ga^\mu A_\mu \big)
\label{bilinear}
\eeq
is the bilinear form of the action (\ref{spi}). 
Here and below we assume but usually do not write explicitly 
the identity matrix $\hat{1}$ in the space of Dirac spinors.

The derivation of (\ref{TrLog}) can be performed by several 
methods. Here we intend to use the heat-kernel approach and 
the  Schwinger-DeWitt technique. For this end we need to 
reduce the problem to the derivation of 
$\,\Ln \Det\,\hat{\cal O}$, where the operator $\,{\cal O}\,$
should have the form
\beq
\hat{{\cal O}} \,=\, 
{\widehat{\Box}} + 2{\hat h}^\mu\na_\mu + {\widehat \Pi}\,.
\label{box op}
\eeq
An obvious way to achieve the desired form of the operator 
is to multiply $\hat{H}$ by an appropriate conjugate 
operator $\hat{H}^*$, 
\beq
\hat{{\cal O}} \,=\, \hat{H}\cdot\hat{H}^{*}
\label{box product}
\eeq
and use the relation
\beq
\Ln \Det\,\hat{H} \,=\,\Ln \Det\,\hat{{\cal O}}
\,-\,\Ln \Det\,\hat{H}^*\,.
\label{TrLog-square}
\eeq

It is clear that one can obtain the desirable form of 
the product by simply choosing $\hat{H}^* = \hat{H}$. 
In this case we can obtain the result (\ref{TrLog}) 
by taking 
\beq
\Ln \Det\,\hat{H} 
\,=\,\frac12\,\Ln \Det\,\hat{H}^2\,.
\label{1-loop-primitive}
\eeq
At the same
time there are many other possible choices. For instance,
the calculations can be performed in the most simple and
economic way if we choose the conjugated operator with the
opposite sign of the mass term, 
\beq
\hat{H}^{*}_1 &=& 
- \,i\,\big(\ga^\mu \na_\mu + im - ie\ga^\mu A_\mu\big)\,.
\label{H1}
\eeq
According to \cite{Guilherme}, the contributions of the 
two operators are identical,
\beq
\Ln \Det\,\hat{H} \,=\,\Ln \Det\,\hat{H}^*_1\,,
\label{Gui}
\eeq
such that we can still use the formula similar to 
(\ref{1-loop-primitive}), 
\beq
\Ln \Det\,\hat{H} 
\,=\,\frac12\,\Ln \Det\,\big(\hat{H}\hat{H}_1^* \big)\,.
\label{1-loop-simple}
\eeq
The expressions for the elements of the operator 
(\ref{box op}) in this case are rather simple, namely 
\beq
{\hat h}_1^\mu &=& - ie A^{\mu}
\cr\cr
{\widehat \Pi}_1 &=& - \frac{1}{4}\,R + m^2 
- \frac{ie}{2}\gamma^\mu \gamma^\nu F_{\mu \nu}
\cr
&-&\,ie\, (\na_{\mu}A^{\mu}) 
\,-\,ie^2\,A^{\mu}A_{\mu}\,.
\label{Ph1}
\eeq
Another possible choice of $\hat{H}^*$ is 
\beq
\hat{H}^{*}_2 &=& -\,i\, \big(\ga^\mu \na_\mu - im \big)\,.
\label{H2star}
\eeq
In this case we have to use the relation
\beq
\Ln \Det\,\hat{H} 
\,=\,\Ln \Det\,\big(\hat{H}\hat{H}_2^* \big)
\,-\,\Ln \Det\, \hat{H}_2^* \,.
\label{1-loop-second}
\eeq
It is obvious that the contribution of $\hat{H}_1^*$
does not depend on $A_\mu$, just because this operator 
does not depend on $A_\mu$. Therefore, for evaluating 
the contribution of $\Ln \Det\,\hat{H}$ we can take 
the $A_\mu$-dependent terms from the contribution of 
the product $\big(\hat{H}\hat{H}_1^* \big)$ with the 
coefficient one, while for the $A_\mu$-independent 
terms the coefficient must be one-half. 

The UV divergences derived within the two schemes, the first 
one with $H_1^{*}$ and the second one with $H_2^{*}$ are 
equal to each other. At the same time there is an essential 
difference between the finite parts of the corresponding 
effective actions derived within the two schemes.
The general discussion of this 
difference (called multiplicative anomaly, 
\cite{MA-0,MA-1}), is discussed in the parallel paper
\cite{multi}. Let us present the technical part of the 
consideration, which are also relevant for our main 
targets, that is violation of conformal symmetry and, 
in general, quantum corrections. 

The elements of the operator $\hat{{\cal O}}$ for the 
case of the operator $\hat{H}_2^{*}$ have the form
\beq
{\hat h}_2^\mu &=& -\frac{ie}{2} 
\ga^{\nu}\ga^{\mu}A_{\nu}\cr\cr
{\widehat \Pi}_2 &=& - \frac{1}{4}\,R + m^2 
+ eM \ga^{\mu}A_{\mu}\,.
\label{Ph2}
\eeq

It proves useful to calculate the first Schwinger-DeWitt
coefficients in an arbitrary space-time dimension $d$. 
For this end we will need the expressions for the 
operators 
\beq
{\widehat P}\,=\,{\widehat \Pi} 
+ \frac{\hat 1}{6}\,R - \na_\mu{\hat h}^\mu
- {\hat h}_\mu{\hat h}^\mu
\label{P}
\eeq
and
\beq
{\widehat S}_{\mu\nu}\,=\,{\widehat R}_{\nu\mu}
- \na_\mu {\hat h}_\nu + \na_\nu {\hat h}_\mu
- {\hat h}_\mu {\hat h}_\nu + {\hat h}_\nu {\hat h}_\mu \,,
\label{S}
\eeq
where ${\widehat R}_{\mu\nu}=\hat{1}\big[\na_\mu , \na_\nu \big]$
is the commutator of the two covariant derivatives acting on the 
corresponding fields. In our case of Dirac spinors space
(see, e.g., \cite{book}), 
\beq
{\widehat R}_{\mu\nu}\,\psi 
\,&=&\, [\na_\mu\,,\,\na_\nu]\,\psi
\,=\,\frac14\,R_{\mu\nu\la\tau}\,\ga^\la\ga^\tau \psi\,,
\cr\cr
\ga^\mu\ga^\nu\na_\mu\na_\nu
\,&=&\,{\widehat{\Box}} - \frac{\hat{1}}{4}\,R\,.
\label{di3}
\eeq 
It is important to remember that formulas (\ref{P})
and  (\ref{S}) do not depend on the dimension $d$, and
the {\it general} expressions for the coincidence limits 
of the Schwinger-DeWitt coefficients do not depend on 
$d$ either. 

The expressions for $\,{\widehat P}\,$ and 
$\,{\widehat S}_{\mu\nu}\,$ for the two calculational 
schemes (with $\hat{H}^{*}_1$ and $\hat{H}^{*}_2$, 
correspondingly) are as follows: 
\beq
{\widehat P}_1 &=&  m^2  - \frac{1}{12}\,R 
- \frac{ie}{2}\gamma^\mu \gamma^\nu F_{\mu \nu}\,,
\label{PS1}
\\
{\widehat S}_{1,\,\mu\nu} &=&  
-\, \frac{1}{4}\,\ga^\al \ga^\be\, R_{\al\be\mu\nu}
\, + \,ie\,F_{\mu \nu} \,;
\nonumber
\eeq
versus
\beq
\hat{P_2} &=& m^2 \,-\, \frac{1}{12}\,R
- \frac{ie}{4}\,\ga^\mu \ga^\nu\, F_{\mu \nu}
+ em \gamma^\mu A_\mu
\cr
&+& \frac{ie}{2} (\na^\mu A_\mu)
- \frac{e^2\,(d-2)}{4}\, A^\nu A_\nu\,,
\label{PS2}
\\
\nonumber
\\
\hat{S}_{2,\,\mu \nu} &=&
- \frac{1}{4}\ga^\al \ga^\be R_{\al \be \mu \nu}
+ \frac{ie}{2}\ga^\al 
\left(\ga_\nu \na_\mu A_\al - \ga_\mu \na_\nu A_\al \right)
\cr
&+&\,\frac{e^2}{4}\,\ga_\al \left(\ga_\mu \ga_\be \ga_\nu 
- \ga_\nu \ga_\be \ga_\mu\right) \,A^\al A^\be \,.
\nonumber
\eeq

The operators (\ref{PS1}) and (\ref{PS2}) enable one to 
calculate the coincidence limits of the first coefficients 
of the Schwinger-DeWitt expansion, by just using the 
known general expressions (see, e.g., \cite{BDW-65,hove}). 

Let us start from the first nontrivial coefficient. 
For the sake of simplicity we shall use the notation
$$
a_k \,=\,
\Tr \lim\limits_{x^\prime\to x}a_k(x,x^\prime)
$$ 
for the functional traces (including space-time 
integrations) of the coincidence limits. 
We know \cite{BDW-65,bavi85} that for the operators 
of the form (\ref{box op}), the first coefficient is 
given by the expression
$$
a_1 \,=\,\sTr \hat{P}\,=\, 
- \int d^dx \sqrt{g}\,\,\tr\,\hat{P}\,,
$$
where the operators $\hat{P}$ are 
given by the expressions (\ref{PS1}) and (\ref{PS2}). 
The symbol $\sTr$ means the functional trace, which
is taken according to Grassmann parity of the corresponding 
functional matrices. Some pedagogical examples, including 
the operators with both bosonic and fermionic blocks can 
be found in \cite{book}.  For the general $d$-dimensional 
case we obtain the expressions
\beq
a_{1,\,1} &=& 
-\,\int d^dx \sqrt{g}\,\Big(4m^2 - \frac{1}{3}\,R
\Big)\,.
\label{a1-1}
\eeq
and 
\beq
a_{1,\,2} &=& 
-\,\int d^dx \sqrt{g}\,\Big\{4m^2 - \frac{1}{3}\,R
\,\,+\,\,  2ie\,(\na^\mu A_\mu)
\cr
&-& (d-2)\,e^2\, A^\mu A_\mu \Big\}\,.
\label{a1-2}
\eeq
The first expression (\ref{a1-1}) is independent of the 
electromagnetic potential and is, therefore, gauge 
invariant. However, the second expression is obviously 
different in both respects, in particular it becomes
gauge invariant only in the $d=2$ case (in what follows we
shall call it $2d$, and exactly the same way for other
dimensions, e.g. $4d$ means $d=4$ etc). 

How can we understand the difference between the results 
coming from the two calculational schemes, based on 
$\hat{H}^{*}_1$ and $\hat{H}^{*}_2$ conjugate operators? 
Let us note that the operation of introducing the conjugate 
operator (\ref{box product}) can be seen in two 
different ways. 

First, we can see it as a change 
of the quantum variables in the path integral 
(\ref{eaction}), that is taking 
$\psi = \hat{H}^{*}\chi$, where $\chi$ is a new 
quantum variable. In this case we have to take 
into account the contribution of the functional 
Jacobian of such transformation. This Jacobian is 
in fact related to another path integral, similar to 
(\ref{eaction}), namely, for the second case,
\beq
\int D\chi  D\bar{\chi}\, \exp 
\Big\{
i \int d^dx\sqrt{g}\, \bar{\chi} \hat{H}_2^*\chi \Big\}\,.
\label{eaction1}
\eeq
This integral does not depend on the electromagnetic 
potential, but only on the metric, and hence it is gauge 
invariant. However, the result of the change of variables 
is quite different, for one meets another functional integral, 
\beq
\int D\chi D\bar{\psi}\, 
\exp \Big\{\,i \int d^dx\sqrt{g}
\,\,\bar{\psi}\,\big(\hat{H}\cdot\hat{H}_2^{*}\big)\,\chi
\Big\}\,.
\label{eaction3}
\eeq
This integral does not possess gauge invariance 
and hence it is not a surprise that the result of 
the change of variables also does not possess it. 

Another possible understanding of the operation 
(\ref{box product}) is through the well-known 
relation 
\beq
\Ln\Det(\hat{H}^{*}\cdot \hat{H}^{*}_2)
\,=\,\Ln\Det \hat{H}^{*} + \Ln\Det \hat{H}^{*}_2\,.
\label{product}
\eeq
The two terms in the {\it r.h.s.} are gauge 
invariant due to the reasons explained above, namely 
the first term is a result of the gauge-invariant 
functional integration and the second term simply
does not transform, just because it does not depend
neither on $A_\mu$ neither on spinor field. Hence
if we find the violation of gauge symmetry in the 
{\it l.h.s.} (as we actually did), this indicates 
the violation of the ``identity'' (\ref{product}). 

The relation similar to (\ref{product}), 
$$
\det(A\cdot B)=\det A \cdot \det B
$$
can be easily proved for the finite-size square matrices 
$A$ and $B$. However, the proof can-not be generalized 
for the differential operators, which have an 
infinite-size matrix representations. In fact, some 
mathematicians and physicists were looking, for a long 
time, for an example when this relation does not hold 
\cite{MA-0,MA-1}, by using the $\ze$-regularization 
technique \cite{zeta}. The phenomenon of such possible 
violation has been called Multiplicative Anomaly. 
However, the results of the mentioned works met 
justified criticism \cite{MA-2,MA-3,MA-4} 
because it is in fact difficult to distinguish 
the effect from the usual renormalization ambiguity. 

From the perspective of Multiplicative Anomaly we can observe, 
that the key relation (\ref{product}) hold in $2d$ and only in 
$2d$. However, this dimension is very special for the $a_{1}$, 
because in this particular dimension the coefficient  $a_{1}$ 
defines the logarithmic divergence of the theory. At this point 
we conclude that, in the case under consideration, the relation 
(\ref{product}) does hold for the logarithmically divergent 
part, but may be violated in the other sectors of the effective 
action. In a moment we shall see that this statement is also 
valid for the next two coefficients.

The general expression for the second Schwinger-DeWitt
term (called also the ``magic'' coefficient) is  
\beq
a_2\,&=&\,\sTr {\hat a}_2(x,x)
\,=\, 
\sTr \Big\{
\frac{{\hat 1}}{180}\,(R_{\mu\nu\al\be}^2
- R_{\al\be}^2+{\Box}R)
\cr
\cr
&+& \frac12\,{\widehat P}^2 + \frac16\,({\Box}{\widehat P})
+ \frac{1}{12}\,{\widehat S}_{\mu\nu}^2\Big\}\,.
\label{a2}
\eeq

Direct calculations using formula (\ref{a2}) and 
the expressions (\ref{PS1}) and (\ref{PS2}) yield the 
following results 
\beq
a_{2,\,1} \, &=& \,\frac{d}{288}\,\int d^d x \sqrt{g}\,  
\big( 48  e^2 F_{\mu\nu}F^{\mu\nu} + R^2 - 24  Rm^2 
\cr
&-& 3 R_{\mu\nu\al\be} R^{\mu\nu\al\be} + 144  m^4 \big).
\label{a2-1}
\eeq
for the first calculational scheme based on $\hat{H}^{*}_1$ 
and
\beq
a_{2,\,2} &=& \frac{d}{288}\,
\int d^d x \sqrt{g} \, \Big\{
24\,d\, e^2 (\na^\mu A^\nu)(\na_\mu A_\nu)  
\cr
&-&96\, e^2 (\na^\mu A^\nu)(\na_\nu A_\mu) 
+ R^2 + 144\, m^4
- 24\, Rm^2 
\cr
&-&\,3\, R_{\mu\nu\al\be} R^{\mu\nu\al\be} 
+ 12(d - 4) (R - 12m^2) A^\mu A_\mu \,e^2 
\cr
&+& 
6(d-2)(d-4)\, A_\mu A^\mu A_\nu A^\nu \,e^4
\, \Big\}
\label{a2-2}
\eeq
for an alternative calculational scheme based on 
$\hat{H}^{*}_2$. It is easy to see that the two 
expressions for $a_2$ presented above 
follow the same pattern as the ones for $a_1$
which we considered earlier. Namely, the expression 
for $a_{2,\,1}$ is gauge invariant independent 
of the space-time dimension $d$ and, in fact, its 
dependence on $d$ is rather simple. On the contrary,
$a_{2,\,2}$ manifest much more complicated 
dependence on $d$. This expression is gauge invariant 
only in the $4d$ case, when it also coincides with 
$a_{2,\,1}$. We know that this coincidence 
is because the $a_2$ defines the logarithmic
divergences in $4d$ and only in $4d$. So, we observe 
that the  logarithmic divergences of the theory are 
scheme independent and that for the divergent sector 
of the theory, in dimensional regularization, the 
relation (\ref{product}) actually takes place. At 
the same time, for any other dimension $d\neq 4$, 
the two expressions are dramatically different. We 
can indicate two aspects of this difference. First, 
the $a_{2,\,2}(d\neq 4)$ expression is not 
gauge invariant. This means the quantum contributions 
depend not only on the transverse part of $A_\mu$, 
but also on the longitudinal part of this vector.  
Let us note that the  $a_{2,\,1}(d\neq 4)$
is perfectly gauge invariant and no longitudinal 
propagation takes place in this case. 

Finally, as a final test of the difference between the
two schemes of deriving the photon propagator, let us
calculate the $a_{3,\,1}$ and $a_{3,\,2}$ coefficients. 
For this end we will need the general expression for 
the $a_3$ coefficient, which
has been derived by Gilkey \cite{Gilkey}, and checked 
by Avramidi \cite{avram-tes} who also derived 
$a_4$ coefficient. This general expression is 
rather bulky and we simply refer an interested reader, 
e.g.,
to the Eq. (2.160) of \cite{avram-tes}. One observation 
is in order. The result of \cite{Gilkey,avram-tes}
correspond to the operator of the more simple form 
\beq
\hat{{\cal O}} \,=\, 
{\widehat{\Box}} + {\widehat \Pi}\,.
\label{box op simp}
\eeq
if compared to our (\ref{box op}). However the 
way to generalize from (\ref{box op simp}) to 
(\ref{box op}) is already known for a long 
time from \cite{hove}, where it was applied for 
the $a_{2}$ coefficient. One has to start
by constructing the generalized covariant 
derivative acting in the fermionic space 
$$
\hat{D}_\mu = \hat{\na}_\mu + \hat{h}_\mu\,.
$$
It is easy to see that the operator (\ref{box op})
can be presented in the form (\ref{box op simp})
with the new derivative and, also, ${\widehat \Pi}$
replaced by ${\widehat P}$. The last thing to do 
is to replace the commutator ${\widehat R}_{\mu\nu}$ 
by ${\widehat S}_{\mu\nu}$, defined in (\ref{S}).
Then the Schwinger-DeWitt technique can be developed 
with the covariant derivative $D_\mu$ instead of 
$\na_\mu$, and eventually gives the result in terms 
of ${\widehat P}$, ${\widehat S}_{\mu\nu}$ and 
curvature tensor. 

Consider the most simple $\hat{H}^{*}_1$-based 
calculational scheme. Using the original approach 
for the operator (\ref{box op simp}) we arrive at 
the already known result of Ref. \cite{Drummond},
\beq
a_{3,\,1}
&=& \frac {d\,e^2}{360} 
\, ( \, 2\,R_{\mu\nu\al\be} \, F^{\mu\nu}F^{\al\be} 
\,-\, 26\,R_{\al\nu} \, F^{\mu\nu} \, F_\mu^{\,\,\,\,\al} 
\cr
&+& 24\,\na_\nu F^{\mu\nu}\, \na_\al F_\mu^{\,\,\,\,\al}
\,+\, 5\,R \, F^{\mu\nu} \, F_{\mu\nu} \,)\,.
\label{a3-1}
\eeq
It is easy to see that this expression is gauge 
invariant independent of the space-time dimension $d$. 
Let us now see what is the situation with the second
choice of the calculational scheme, based on the 
operator $\hat{H}^{*}_2$. In this case one has to 
use the approach described above, that is to deal 
with the operator (\ref{box op}) and use generalized
covariant derivative $D_\mu$. After an involved 
algebra we arrive at the expression
\begin{widetext}
\beq
a_{3,\,2}\Big|_{AA} 
&=& 
\frac{de^2}{2880} \Big\{ 120(\na A)\Box(\na A)
- 60 F_{\mu\nu}\Box F^{\mu\nu}
- 24\na_\nu F^{\mu\nu}\na^\al F_{\mu\al}
\,+\,
 24(\Box A^\al)\big[(d-3)(\Box A_\al) 
+ 2\na_\al(\na A)\big]
\cr
\cr
&-&
24(\na_\al\na_\mu A_\be)\big[(\na^\be\na^\mu A^\al) 
- (\na^\al\na^\mu A^\be)\big] 
\,+\,
 A^\mu A_\mu 
\big[(18-7d)R^2_{\mu\nu\al\be}
-8(9-d)R^2_{\mu\nu}-6(5-d)R^2\big]
\cr
\cr 
&+& 8R_{\mu\nu\al\be}\big[
4 (\na^\al A^\nu)(\na^\mu A^\be)
- 8 F^{\mu\nu}F^{\al\be}
\,-\,
3(d-4) (\na^\mu A^\al)(\na^\nu A^\be)
- R^{\la\nu\al\be}A^\mu A_\la
+ 10 R^{\mu\be}  A^\al A^\nu
\big]
\cr
\cr
&+&  16 R_{\mu\nu} 
\big[
10 (\na A)(\na^\mu A^\nu)
\,+\,
(\na^\al A^\mu)(5\na_\al A^\nu
- 2\na^\nu A_\al)
\,-\, 
(d-5)(\na^\mu A^\al)(\na^\nu A_\al)
-2 R^\mu_{\,\,\,\al}\,A^\al A^\nu
\big]
\cr
\cr
&+&  10R \big[
2(d-5)(\na_\mu A_\nu)(\na^\mu A^\nu)
- 2 (\na A)^2
+ 3 F_{\mu\nu} F^{\mu\nu}
+ 2 R_{\mu\nu} A^{\mu}A^{\nu}
\big]
\,-\, 24(\na^\nu R)
[(\na_\nu A^\al A_\al) - (\na_\al A^\al A_\nu)]
\cr
\cr
&-&  12(d-2)A^\al A_\al \Box R 
- 48(\na^\al R_{\mu\nu\al\be})
(\na^\nu A^\be A^\mu) \Big\}\,.
\label{a3-2}
\eeq
\end{widetext}
The last expression is rather cumbersome and difficult
to deal with. Apparently it is not gauge invariant and 
is different from (\ref{a3-1}). In order to complete the 
analysis, we need to answer two questions: 
\ {\large\it (i)} How we
can prove that the two expressions are distinct or 
identical in one or another particular dimension? 
\ \ {\large\it (ii)}  
Is it true that the two expressions (\ref{a3-1}) 
and (\ref{a3-2}) are distinct, in one or another 
particular dimension? 
In case of $a_3$ we are mainly interested in 
the $6d$ case, where one can hope to find the 
gauge invariance of (\ref{a3-2}) and also the
equivalence between (\ref{a3-1}) and (\ref{a3-2}).   

Let us start by making the most simple test. Consider
the propagation of the purely transverse part of $A_\mu$,
that means to collect the terms which contain the terms
$A_\mu \Box^2 A^\mu$ in both expressions. It is important 
that the transverse terms can-not be affected by the 
possible violation of gauge invariance in the 
expression (\ref{a3-2}), so this particular test 
is independent from of rest. The relatively simple 
calculation shows that the two propagators are 
identical in $6d$ and only in $6d$. This solves 
one-half of the item {\large\it (ii)}, namely we 
learn that out of $6d$ the two expressions (\ref{a3-1}) 
and (\ref{a3-2}) are different. 

We could not precisely prove that the two expressions are 
identical in $6d$ for an arbitrary metric background. It is 
likely, however, that this is the case. The arguments in 
favor of this conclusion come from the analysis of a
particular simple metrics. 
Let us consider the case of de Sitter space-time, where one
can actually check the equality of Eqs. (\ref{a3-1})  
and (\ref{a3-2}). Even in this case it is not easy to work 
with (\ref{a3-2}), but one can perform some qualitative 
analysis of the expression for $a_{3,\,2}$. First 
of all, by direct inspection we can verify that the terms 
without scalar curvature are the same in both expressions. 

As a next step, consider the terms proportional to
the scalar curvature, which can be symbolically identified 
as $RFF$. In order to analyze these terms, one has to take 
into account also the terms proportional to the derivatives 
of the curvature tensor components in (\ref{a3-2}). For 
definiteness, we call them DR-terms. If one integrates 
these terms by parts, it turns out that some of them do
contribute to the $R^2A^2$-structures. Since these terms can 
be written as total derivative terms in de Sitter space-time, 
we can introduce them in $a_3^{(2)}$ with an arbitrary 
coefficient. To make the correct choice of this coefficient,
one has to additionally evaluate those terms which are 
proportional to $R^2A^2$ and choose the coefficient of 
the DR-terms such that they do cancel the $R^2A^2$ terms 
in (\ref{a3-2}). After that, the terms proportional to 
$RFF$, generated by integration by parts of the DR-terms, 
sum with the ones which come from (\ref{a3-2}) and 
finally turn out to be the same as (\ref{a3-1}). We do 
not include the corresponding formulas here, because they 
are rather boring. 
\vskip 2mm

Finally, we have a strong argument to believe that 
qualitatively the situation with $a_3$ is the 
same as with the $a_1$ and $a_2$
coefficients. All these coefficients are universal 
(scheme independent) in the dimensions $2,\,4$ and 
$6$, correspondingly. And, at the same time, they 
are essentially scheme dependent in other dimensions. 
It is easy to see that this fact implies that, in 
any dimension, the logarithmic divergences are 
scheme independent and gauge-invariant. At the same
time, quartic and quadratic divergences and the 
finite contributions are scheme dependent and, in 
case of the $\hat{H}^{*}_2$-based calculational 
scheme, they are not gauge-invariant. Of course, 
this feature does not contradict the gauge-invariant 
renormalizability of the theory, but 
it makes the finite results scheme dependent. 
One has to remember that the effective action,
in any particular dimension, is a sum of the 
series of the terms with ${\hat a}_0$, ${\hat a}_1$, 
${\hat a}_2$, ${\hat a}_3$ etc. Therefore, we can 
conclude that the effective action is scheme dependent 
in any particular dimension. 

The scheme dependence which we have explored 
(multiplicative anomaly), should be seen as a particular 
case of the nontrivial reaction of the effective action 
of quantum fields to the change of variables. The fact
that such dependence can be relevant, has been known for 
a long 
time (see, e.g., \cite{tyutin}). It is interesting to see 
what is the consequence of this fact for the sum of the 
Schwinger-DeWitt series. We perform the corresponding 
calculation in the next section.

\section{Derivation of the one-loop formfactors}

The effective action (\ref{TrLog}) can be expressed by 
means of the proper time integral, involving the 
heat-kernel $K(s)$ of the operator ${\cal O}$, 
\beq
\bar{\Ga}^{1} = -\frac{1}{2}\int_0^\infty\,\frac{ds}{s}
\,\sTr K(s)\,.
\label{heat}
\eeq
Let us derive this expression using the method developed
previously in \cite{apco,fervi} (see also \cite{Poimpo} 
for the general review). According to Ref. \cite{apco}, 
we can perform calculations either by using Feynman 
diagrams or through the heat-kernel solution, which 
was obtained in \cite{bavi90,Avramidi89}. The result 
of the two methods are going to be the same, but the 
covariant heat-kernel solution is much more simple, 
so we will use this approach. 
 
The expression (\ref{heat}) can be expanded into 
the powers of the field strengths (curvatures), 
$\,R_{\mu\nu\al\be}$, $\,\hat{S}_{\mu\nu}\,$ 
and $\,\hat{P}$. Up to the second order in the 
curvatures the expansion has the form \cite{bavi90}
\beq
\Tr K(s) & = & \frac{\mu^{4-2\om}}{(4\pi s)^\om}
\,\int d^{2\om -4}x\,\sqrt{g}
\,\,e^{-sM^2}\tr \{\, \hat{1}+ s\hat{P}
\cr
&+&\, s^2\,\big[\, 
R_{\mu\nu}\,f_1(-s\Box ) \,R^{\mu\nu} 
+ R\,f_2(-s\Box )\,R 
\cr
&+& \hat{P}\,f_3(-s\Box )\,R 
+ \hat{P}\,f_4(-s\Box )\,\hat{P} 
\cr
&+& \hat{S}_{\mu\nu}\,f_5(-s\Box )\,\hat{S}^{\mu\nu}
\big]\,\}\,.
\label{heat 2}
\eeq
where the expressions for $\hat{P}$ and $\hat{S}_{\mu\nu}$
are defined in (\ref{P}) and (\ref{S}), \ $\om$ \ is the
dimensional regularization parameter, $\mu$ is an arbitrary 
renormalization parameter with the dimension of mass and 
the functions $f_i$ are given by
\\
\beq
f_1(\ta) &=& \frac{f(\ta)-1+\ta /6}{\ta^2}\,,
\cr
f_2(\ta) &=& \frac{f(\ta)}{288}+\frac{f(\ta)-1}{24\ta}-
\frac{f(\ta)-1+\ta /6}{8\ta^2}\,,
\nonumber
\\
\nonumber
\\
f_3(\ta) &=& \frac{f(\ta)}{12}+\frac{f(\ta)-1}{2\ta}
\,, \;\;\;\;\;\;
f_4(\ta) = \frac{f(\ta)}{2}\, , 
\cr 
f_5(\ta) &=& \frac{1 - f(\ta)}{2\ta}\,,
\label{fk}
\eeq
where
\beq
f(\ta) &=& \int_0^1 d\al\, e^{\al (1-\al)\ta}\,, 
\;\;\;\;\; \ta = -s\Box\,.
\nonumber
\eeq
\vskip 2mm

After some algebra, we arrive at the integral representation
for the effective action up to the second order in 
$\,{\hat P}$, $\,{\hat S}_{\mu\nu}\,$ and $\,R_{\mu\nu}$,
in dimensional regularization,  
\begin{widetext}
\beq
\bar{\Ga}^{1}\,&=&\, -\frac{1}{2(4\pi)^2}
\,\int d^{4}x\sqrt{g}\,
\left(\frac{m^2}{4\pi\mu^2}\right)^{\om-2}
\int_0^\infty\,dt \,e^{-t}\,
\sum_{k=1}^{5}\,\Big\{\, 
l^{FF}_k\, e^2 F_{\mu\nu}\,M_k\,F^{\mu\nu}
\,+\,
l^{*\,DA}_k\, e^2 \na_\mu A^\mu\,M_k\,\na_\nu A^\nu
\cr
\cr
&+&
l^{DA}_k\, e^2 \na_\mu A^\nu M_k\,\na_\nu A^\mu
\,+\,
\la_k\, e^2 R_{\mu\nu} M_k\,A^\nu A^\mu
\,+\,
\la^*_k\, e^2 A^\nu A^\mu M_k\, R_{\mu\nu}
\,+\,
l^{AR}_k\, e^2 A_\al A^\al  M_k\, R
\cr
\cr 
&+&
l^{RA}_k\, e^2 R  M_k\, A_\al A^\al
\,+\,
l^*_k R_{\mu\nu}\,M_k\,R^{\mu\nu} 
\,+\,
l_k R\,M_k\,R 
\,\Big\}\, ,
\label{EA}
\eeq
where 
\beq
M_1 = \frac{f(tu)}{t^{\om -1}}\, , \;\;\; M_2 
= \frac{f(tu)}{t^\om u}\, , \;\;\;
M_3 = \frac{f(tu)}{t^{\om +1}u^2}\, ,\;\;\; M_4 
= \frac{1}{t^\om u}\, ,\;\;\; 
M_5 = \frac{1}{t^{\om +1}u^2}\, , \nonumber
\eeq
\end{widetext}
and the nonzero coefficients for $\hat{H^{*}_1}$ have the form
\beq
l_2 &=& -\frac{1}{16} \qquad l_3 = -\frac{1}{8} \,,\qquad
l_4 = \frac{1}{24} \qquad l_5 = \frac{1}{8}\,,
\cr \cr
l^*_2 &=& \frac{1}{4} \,,\qquad l^*_3 = 1 \,,\qquad
l^*_4 = -\frac{1}{12} \,,\qquad l^*_5 = -1\,,
\cr \cr
l^{FF_{(1)}}_1 &=& 1\,,\quad 
l^{FF_{(1)}}_2 = 2\,, \quad 
l^{FF_{(1)}}_4 = -2\,.
\nonumber
\eeq
Indeed, the coefficients $\,l_k\,$ and $\,l^*_k\,$ are 
the same as for the free fermion field \cite{fervi}, while 
the last set of coefficients is due to the interaction
with $A_\mu$ and do not have analogues in the free field 
case.
For $\hat{H^{*}_2}$, we have
\beq
l^{FF_{(2)}}_1 &=& \frac{1}{4}\,, 
\qquad l^{FF_{(2)}}_2 = -\frac{1}{2}\,, 
\qquad l^{FF_{(2)}}_4 = \frac{1}{2}\,,
\cr \cr
l^{*\,DA}_1 &=& -\frac{1}{2} \,, \qquad 
l^{*\,DA}_2 = -1 \,, \qquad 
l^{*\,DA}_4 = 1 \,,
\cr \cr
l^{DA}_2 &=& -2 \,, \qquad 
l^{DA}_4 = 2 \,,
\cr \cr
\la_2 &=& -1 \,, \qquad 
\la_4 = 1 \,, \qquad
\la^*_2 = -1 \,, \qquad 
\la^*_4 = 1 \,, 
\cr \cr
l^{AR}_1 &=& -\frac{1}{6} \,, \qquad 
l^{AR}_2 = -1 \,, \qquad 
l^{RA}_4 = 1 \,,
\cr \cr
l^{RA}_1 &=& \frac{1}{6} \,, \qquad 
l^{RA}_2 = 1 \,, \qquad 
l^{RA}_4 = -1 \,.
\eeq
\vskip 2mm

As far as the purely gravitational terms were calculated
in \cite{fervi}, here we are mainly interested in the 
one-loop corrections to the term $F^2_{\mu\nu}$. Let us first 
perform all the calculations for the operator $\hat{H^{*}_1}$. 
Substitute (\ref{PS2}) 
into (\ref{heat 2}), then the result into (\ref{heat}), 
and take into account only the terms with $f_4$ and $f_5$. 
Using the notations \cite{apco},
$$
\,t=sM^2\;,\quad u=\frac{\ta}{t} = -\frac{\Box}{M^2}\,,
$$ 
\beq
Y\,=\,1-\frac{1}{a}\ln\,\Big(\frac{2+a}{2-a}\Big)\,,
\qquad a^2 = \frac{4\Box}{\Box - 4M^2}
\label{Aa}
\eeq
and the known results for the integrals \cite{fervi},
\beq
\Big(\frac{M^2}{4\pi \mu^2}\Big)^{\om - 2}
\int_0^\infty dt \; e^{-t}\, \frac{f(tu)}{t^\om u} 
&=& 
\cr 
\Big[
\left(\frac{1}{12}-\frac{1}{a^2}\right)\,\left(
\frac{1}{\ep} + 1 \right) - \frac{4A}{3a^2}
&+& \frac{1}{18}\Big] + {\cal O}(2-\om)\, , 
\cr
\cr
\cr
\Big(\frac{M^2}{4\pi \mu^2}\Big)^{\om - 2}
\int_0^\infty dt \; e^{-t}\, \frac{1}{t^\om u} 
&=& 
\cr
\Big[ \frac{a^2-4}{4\,a^2}
\,\left(\,\frac{1}{\ep} + 1\right)\Big] + {\cal O}(2-\om )\, , 
\cr
\cr
\cr
\Big(\frac{M^2}{4\pi \mu^2}\Big)^{\om - 2}
\int_0^\infty dt \; e^{-t}\, t^{1 - \om} f(ut) 
&=& 
\cr
\,\left(\,\frac{1}{\ep} + 2A\right) + {\cal O}(2-\om )\, , 
\nonumber
\eeq
where we also denoted
$$
\frac{1}{\ep}=\frac{1}{2-\om }
+\ln \Big(\frac{4\pi \mu^2}{M^2}\Big) \,.
$$

After some algebra, we arrive at the explicit expression 
for the one-loop correction to the classical 
$F^2_{\mu\nu}$-term, 
\beq
{\bar \Ga}^{(1)}_{\sim F^2_{\mu\nu}}
= - \frac{e^2}{2(4\pi)^2}\int d^4x \sqrt{g}
F_{\mu\nu} 
\Big[\frac{2}{3\,\ep}+k^{FF}_1(a)\Big] F^{\mu\nu},
\label{FF1}
\eeq
where the electromagnetic formfactor has the form
\footnote{We note that this result differs from the 
similar expression previously derived in \cite{Milton}. 
Unfortunately 
we can not indicate the source of the difference, 
but there are strong reasons to believe that it is 
not a scheme dependence which is discussed here and 
also in \cite{multi}. In particular, the one-loop 
result of \cite{Milton} involves double logarithmic 
terms which we did not meet. Because of this, it 
does not reproduce a known expression in the massless
limit.}
\beq
k^{FF}_1(a)=
Y\Big(2-\frac{8}{3a^2}\Big)-\frac{2}{9}\,.
\label{K1}
\eeq

A simple way to check our result is to compare it to 
the expressions for the Schwinger-DeWitt coefficients 
derived in the previous section. Let us remember that 
the calculation of the formfactor presented above has 
been performed in $4d$. There is nothing that can be 
compared to $a_1$, just because 
the quadratic divergences vanish when the dimensional 
regularization is used. The comparison to $a_2$ is 
absolutely successful and complete, because here we
meet two perfect correspondences. First, it is easy 
to see that the coefficient of the pole term of 
(\ref{FF1}) is exactly the $a_2$ from 
(\ref{a2-1}) at $4d$. Second, it is easy to check 
that the UV limit $a\to 0$ of the formfactor (\ref{K1}) 
perfectly corresponds to the coefficient of the pole 
term of (\ref{FF1}). 

Furthermore, we can partially compare our formfactor 
(\ref{K1}) to the expression for the $a_{3,\,1}$ 
in (\ref{a3-1}). Unfortunately, the complete verification 
is not possible because certain part of (\ref{a3-1})
correspond to the next, third, order in the heat-kernel 
solution, which we do not use here. So, we can compare 
only the $\,F\na\na F$-type terms. 
If we expand the corresponding part of $k^{FF}_1(a)$ 
into power series of the parameter $a$, in the IR limit, 
where 
\beq
a^2 \,\sim \,-\, 4\,\frac{\Box}{M^2}\,,
\label{assimptotic}
\eeq 
use Maxwell equation
\beq
\na_\ro F_{\mu\nu} + \na_\mu F_{\nu\ro} 
+ \na_\nu F_{\ro\mu} = 0 
\label{FFsim}
\eeq
and integrate by parts, we obtain the corresponding term in 
the effective action (\ref{a3-1}), 
\beq
-\, \frac{1}{120 \pi^2} 
\na_\mu F^{\mu\nu} \na_\ro F^\ro_{\,\,\,\,\, \nu}\,,
\label{1-120}
\eeq
which is actually well-known from \cite{Drummond}.
\vskip 4mm

Consider the second calculational scheme, based on 
the operator $\hat{H}^*_2$, defined in Eq. (\ref{H2star}). 
The calculations are a bit more involved, but are in general  
analogous to the previous case. Finally, we arrive at the 
following result 
\beq
{\bar \Ga}^{(1)}_{\sim A^2}
&=& - \,\frac{e^2}{2(4\pi)^2}\,\int d^4x \,\sqrt{g}\,\,\Big\{
F_{\mu\nu} 
\,\Big[\frac{2}{3\,\ep}+k^{FF}_2(a)\Big] F^{\mu\nu} 
\cr
&+&
\na_\mu A^\mu \,\Big[\,  Y \,\Big( \frac{8}{3a^2} -2 \Big) \,+\, 
\frac {2}{3} \,\Big] \na_\nu A^\nu 
\cr
&+&\,
\na_\mu A^\nu \,\Big( \, \frac{16Y}{3a^2}  
\,\Big) \na_\nu A^\mu
\,+\,
R_{\mu\nu} \,\Big( \frac{8Y}{3a^2} \,\Big) A^\nu A^\mu 
\cr 
&+&\,
A^\nu A^\mu \,\Big( \, \frac{8Y}{3a^2} \,\Big) R_{\mu\nu} 
\,+\,
A_\al A^\al 
\,\Big[\,  Y \,\Big( \frac{1}{3} - 
\frac{4}{3a^2} \Big) \,\Big]\, R 
\cr
&+&
R\,\Big[\,  Y \,\Big( \frac{4}{3a^2} - 
\frac{1}{3} \Big) \,\Big]\, A_\al A^\al
\Big\}\,,
\label{FF2}
\eeq
\beq
k^{FF}_2(a)=
Y\Big(1+\frac{4}{3a^2}\Big)+\frac{1}{9}\,.
\label{K2}
\eeq
Once again, the UV test and the comparison with the $a_{3,\,2}$
are perfectly successful. However, the low-energy expression is 
not (\ref{1-120}) anymore in this case, instead we have 
\beq
-\, \frac{1}{160 \pi^2} 
\na_\mu F^{\mu\nu} \na_\ro F^\ro_{\,\,\,\,\, \nu}\,.
\label{1-80}
\eeq
As we will see in Sect. 6, this divergence between 
(\ref{1-120}) and (\ref{1-80}) results also in the 
ambiguity in the decoupling theorem. 
\vskip 2mm

It is possible to make an interesting verification 
of the results for the two formfactors $k^{FF}_1(a)$ 
and $k^{FF}_2(a)$ using the coefficients $\hat a_{3,\,1}$ 
and $\hat a_{3,\,2}$. If we expand (\ref{K1}) to the 
first order in operator $\Box$, we have
\beq
k^{FF}_1(a)=
Y\Big(2-\frac{8}{3a^2}\Big)-\frac{2}{9} \simeq 
-\frac{2}{15}\frac{\Box}{M^2}\,,
\eeq
that gives us a term in $\bar \Ga^{(1)}$ 
which has the form
\beq
\bar \Ga^{(1)} \,\sim \, \frac{e^2}{15\,(4\pi)^2}
\int d^{4} x \sqrt{g} 
\, F^{\mu\nu} \, \frac{\Box}{M^2} \, F_{\mu\nu}\,.
\label {pass1}
\eeq
Using (\ref{FFsim}) in (\ref{pass1}), we can see that
\beq
\bar \Ga^{(1)} &\sim& - \,
\frac{2e^2}{15\,(4\pi)^2}\,\int d^{4} x \sqrt{g}  
\,\na_\nu F^{\mu\nu}\, \na_\al F_\mu^{\,\,\,\,\al} 
\cr &=&
-\,\frac{24\,e^2}{180\,(4\pi)^2 \,M^2}
\int d^{4} x \sqrt{g}
\na_\nu F^{\mu\nu} \na_\al F_\mu^{\,\,\,\,\al}
\label {pass}
\eeq
and that is the same contribution given to $\bar \Ga^{(1)}$ 
in $4d$ by the third term of Eq. (\ref{a3-1}), as it should be. 
To the form factor $k^{FF}_2(a)$, we find
\beq
\bar \Ga^{(1)} \sim \,-\,
\frac{e^2}{10\,(4\pi)^2}\,\int d^{4} x \sqrt{g} \, 
A^\mu \, \frac{\Box^2}{M^2} \, A_\mu\,.
\label {pass2}
\eeq
Now if we take the contribution of (\ref{a3-2}) to 
$\bar \Ga^{(1)}$ in $4d$ the coefficient proportional to 
$A^\mu \,\Box^2 \, A_\mu$ is the same as (\ref{pass2}). 
It is important to note that the coefficients of the terms 
$A^\mu \,\Box^2 \, A_\mu$ in Eqs. (\ref{pass}) and 
(\ref{pass2}) are different in $4d$. They are equal only 
in $6d$ as it was explained in Sect. $3$.

\section{Massless limit and trace anomaly}

In order to better understand the physical sense of 
the result (\ref{FF1}) and also its relation with the 
MS-based anomaly-induced action (\ref{em-1-loop}), let 
us consider the high-energy limit. As was already 
explained in the previous sections, this can be done 
by taking the vanishing mass limit. On the other hand, 
this limit helps to establish the relation between 
the effective action (\ref{FF1}) and the conformal 
 anomaly (\ref{mainequation}).

Consider the high-energy limit, when the mass 
of the quantum field (i.e. of electron) is negligible. 
Taking the limit $a\to 2$ in the expression (\ref{FF1})
with either one of the two available form factors, we 
arrive at the following leading-log behavior of the 
electromagnetic sector:   
\beq
\tilde{\be}
\,F^{\mu\nu}\,\ln \Big(\frac{\Box}{\mu^2}\Big)\,F_{\mu\nu}\,.
\label{formQED}
\eeq
Similar asymptotic behavior takes place also in the 
gravitational sector of the theory. For instance, the 
Weyl term has similar form factor \cite{apco,fervi},
\beq
\be_1\,C^{\mu\nu\al\be}\,\ln \Big(\frac{\Box}{\mu^2}
\Big)\,C_{\mu\nu\al\be}\,.
\label{formWeyl}
\eeq
Let us note that the same asymptotic behavior can be 
recovered in the Minimal Subtraction - based scheme of 
renormalization \cite{buwolf,book} (see also \cite{Maroto} 
for an alternative consideration), which is completely 
reliable in the massless case. 

The expressions (\ref{formQED}) 
and (\ref{formWeyl}) are sufficient to derive the 
corresponding parts of the conformal anomaly, even 
in case of a local conformal symmetry. For this end,
let us apply the conformal parametrization of the 
metric (\ref{vec trans}) and the differential 
relation (\ref{deriv}). Consider the case of  
(\ref{formQED}) as an example, (\ref{formWeyl}) 
is completely analogous. If we replace the 
parametrization  (\ref{vec trans}) into (\ref{formQED}),
the only place where the $\si$ field shows up is the 
$\Box$. This operator becomes 
\beq
\Box = e^{-2\si}\,\Big[\boxdot + {\cal O}(\pa \si)\Big]\,,
\label{box conf}
\eeq
where we denote ${\boxdot}$ the d'Alembertian 
operator constructed with the metric ${\bar g}_{\mu\nu}$,
$$
{\boxdot}\,=\,
{\bar g}^{\mu\nu} {\bar \na}_\mu {\bar \na}_\mu\nu\,.
$$

The explicit form of the terms ${\cal O}(\pa \si)$ 
is in fact irrelevant for us. When we apply  (\ref{deriv}),
only the first term in the bracket (\ref{box conf}) is 
important, because the other terms vanish after we 
set $\,\si \to 0$. Of course, the logarithmic dependence 
makes
\beq
\ln \frac{\Box}{\mu^2} = - 2\si + 
 \ln \frac{{\boxdot} + {\cal O}(\pa \si)}{\mu^2}\,.
\label{log box}
\eeq
Finally, after applying  (\ref{deriv}) we arrive at
\beq
<T^\mu_\mu>_{em}\,=\,\tilde{\be} F_{\mu\nu}^2 
\label{Te}
\eeq
in the electromagnetic sector. In a similar way one 
can obtain the $\om C^2$-term in the general formula
for anomaly, Eq. (\ref{mainequation}). One has to 
note that the $\Box R$ term and the Gauss-Bonnet 
term can be also derived from the 
local and nonlocal finite parts of effective action 
\cite{bavi90,bavi3}. At the same time, the $\Box R$
part is a subject of an important ambiguity. The 
origin and mechanism for this ambiguity has been 
explained recently in \cite{AGS} (see also 
\cite{PoS-Conform} for the review). 

\section{
Renormalization group, low-energy limit and decoupling}

The two form factors (\ref{K1}), (\ref{K2}) contain 
all necessary information about the scale dependence of 
the coupling parameter $e$ at the one-loop level, within 
the corresponding calculational schemes. In the present 
section we shall mainly restrict our attention to the 
flat space case and discuss this dependence in detail.
Namely, our task here is to calculate the ``physical''
beta-functions in the momentum-subtraction renormalization 
scheme and look at their UV and IR limits.  

In the $\overline{\rm MS}$ scheme the $\be$-function
of the effective charge $e$ is defined as 
\beq
\be_e(\overline{\rm MS})
\,=\,\lim_{n\to 4}\,\mu\,\frac{de}{d\mu}
\,=\,\frac{4\,e^3}{3\,(4\pi)^2}\,.
\label{beta}
\eeq

The derivation of the $\be$-functions in the mass-dependent 
scheme has been described, e.g. in \cite{Ramond,manohar}. 
Starting 
from the polarization operator, one has to subtract the 
counterterm at the momentum $p^2=M^2$, where $M$ is the 
renormalization point. Then, the momentum-subtraction 
$\be$-function is defined as
\beq
\be_e = \lim_{n\to 4}\,M\,\frac{de}{dM}\,.
\label{beta-mass-M}
\eeq

Mathematically, this is equivalent to taking the derivative 
(we also write the same operation in terms of $a$)
\beq
-e\,p\,\frac{d}{dp}\,=\,e\, (4-a^2)\frac{a}{4}\frac{d}{da}
\eeq
of the form factors in the polarization operator. 
If we apply this procedure to the form factor $k^{FF}_1(a)$ of 
the $F_{\mu\nu}^2$-term, the expression for the $\be$-function
in a mass-dependent scheme is
\beq
\be_e^1\,&=&\, \frac{e^3}{6a^3\,(4\pi)^2}\, 
\Big\{ 20 a^3 - 48 a   
\cr
&+& 3(a^2-4)^2 \,\ln \Big( \frac{2+a}{2-a} \Big)
\Big\}\,,
\label{beta1}
\eeq
that is the general result for the one-loop $\be$-function
valid at any scale \footnote{This expression is essentially 
different from the one presented earlier in \cite{Su}. 
Since this publication does not contain sufficient technical 
details, we could not find the source of this divergence.
At the same time we note that the expression of \cite{Su}
does not produce the well-known large-mass limit 
(Appelquist and Carazzone theorem) which we successfully 
derive below in (\ref{beta1-IR}) from our general formula  
(\ref{beta1}).}.

As the special cases we meet the UV limit $p^2 \gg m^2$, or
$a \rightarrow 2$,
\beq
\be_e^{1\,\,UV} \,=\,\frac{4\,e^3}{3\,(4\pi)^2}\,
+ \,{\cal O}\Big(\frac{m^2}{p^2}\Big)\,,
\label{beta1-UV}
\eeq
that is nothing else but the $\overline{\rm MS}$ scheme
result (\ref{beta}) plus small correction. In the IR 
regime, however, when $p^2 \ll m^2$, the result is quite
different, and moreover depends on the calculational 
scheme. For the first case $\hat{H}^*_1$, we have 
\beq
\be_e^{1\,\,IR} \,=\, \frac{e^3}{(4\pi)^2}\,\cdot\,
\,\frac{4\,M^2}{15\,m^2} \,\,
+ \,\,{\cal O}\Big(\frac{M^4}{m^4}\Big)\,.
\label{beta1-IR}
\eeq
This is exactly the standard form of the decoupling 
theorem \cite{AC}. 

Similar calculations starting from the form factor 
$k^{FF}_2(a)$ give 
\beq
\be_e^2\,&=&\, \frac{e^3}{12a^3\,(4\pi)^2}\, 
\,\Big\{   4a (12+a^2) \,
\cr
&-&3(a^4-16)\,
\ln \Big(\frac{2-a}{2+a}\Big) \Big\}\,.
\label{beta2}
\eeq

In the UV limit $p^2 \gg m^2$, the above $\be$-function is
in agreement with the standard result (\ref{beta1-UV}),
while in the IR limit $p^2 \ll m^2$ we obtain
\beq
\be_e^{2\,\,IR} \,=\, \frac{e^3}{5\,(4\pi)^2}\,\cdot\,
\frac{M^2}{m^2} \,\,
+ \,\,{\cal O}\Big(\frac{M^4}{m^4}\Big)\,.
\label{beta2-IR}
\eeq

As we can see from the expressions (\ref{beta1-IR}) and 
(\ref{beta2-IR}), there is a slight difference in how
$\be$-functions go to zero in the IR limit. To discuss the 
physical sense of this fact, let us take the difference 
between the two form factors (\ref{K1}) and (\ref{K2})
\beq
\Delta F^{FF}\,=\,
k^{FF}_1(a)\,-\,k^{FF}_2(a)\,=\,
A \Big( 1-\frac{4}{a^2} \Big) -\frac{1}{3}
\eeq
and expand $\Delta F^{FF}$ in power series of $a$. In the 
IR limit we have $a^2 \sim p^2\, / \,M^2=\,-\, \Box / M^2$, 
so
\beq
\Delta F^{FF}\,=\, \frac{1}{30} \cdot
\frac{\Box}{M^2} +{\cal O}(p^3/M^3)\,.
\eeq
This is exactly the difference in the form factors which 
caused the ambiguity (calculational scheme dependence) 
in the decoupling theorem. In order to evaluate the 
source of this difference in the effective action, 
we should consider what would be the new terms 
in the equation of motion, generated by the term
\beq
F^{\mu\nu} \Big(\frac{1}{30} \cdot \frac{\Box}{M^2}\Big)
F_{\mu\nu}
\label{decouple}
\eeq  
As this term is proportional to the operator $\Box$, if we 
are working in flat spaces, we use Eq. (\ref{FFsim}) 
to obtain a term proportional to $(\na_\mu F^{\mu\nu})^2$. 
This term will not influence the equations of motion in 
flat space in the ${\cal O} (e^2)$ approximation \cite{Drummond}. 
However, as we will discuss later on, the situation can be
different in curved space.

\section{Running charge}

Let us now consider the running of the electromagnetic 
charge. The difference in the expressions for the form factors 
$k_1^{FF}(a)$ and $k_2^{FF}(a)$ and eventually in the expressions 
for the $\beta$-functions (\ref{beta1}) and (\ref{beta2}), means 
that the effective charge (i.e., the running charge) may have 
different behavior for the different calculational schemes 
(that means, for the ones based on ${\hat H}_1^*$ and 
${\hat H}_2^*$ operators). Also, we expect that the running 
within the Momentum Subtraction scheme will be distinct
from the one within the Minimal Subtraction scheme, especially 
in the low-energy region. Let us check out what the real 
situation is.

To investigate the renormalization of the corresponding 
quantities, in the physical scheme, let us apply the operator 
$-epd/dp$ to $k_{1,2}^{FF}(a)$ with $\Box$ traded for $-p^2$. 
In doing so, we find the expressions for the $\beta$-functions, 
which can be conveniently presented as 
\beq
-\frac{(4\pi)^2}{e^3}\beta_1 &=& p\frac{d}{dp}k_1^{FF}(a) 
= \frac{d}{dt}k_1^{FF}(a)
\quad 
{\rm and}
\cr
-\frac{(4\pi)^2}{e^3}\beta_2 &=&  p\frac{d}{dp}k_2^{FF}(a) = 
\frac{d}{dt}k_2^{FF}(a)\,.
\eeq
Here $t$ is a dimensionless parameter defined by 
$t = \ln (p/m)$.

The UV limit is achieved for $p \gg m$, or equivalently $t \gg 1$,
while in the IR limit we meet inverse relations $p \ll m$ and 
$t \ll 1$. With the help of the {\it Mathematica}
computer software, one can calculate explicitly the above beta 
functions and integrate them. For a clear illustration, we plot
the beta functions for both cases as functions of the parameter 
$a$ (see Fig.~1).


\vskip 5mm
\begin{figure}
\includegraphics[width=0.5\textwidth]{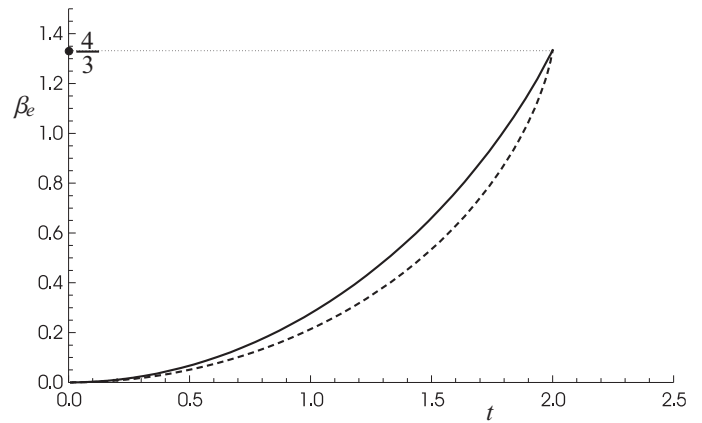}
\begin{quotation}
\caption{The beta functions corresponding to form factors $k_1$ 
(solid line) and $k_2$ (dashed line). The vertical axis is 
drawn in units of $e^3/(4\pi)^2$.}
\end{quotation}
\label{Figura1N}
\end{figure}

The integration of the renormalization equation corresponding 
to $\beta_1$ can be performed by using the {\it Mathematica} 
software such that the integral curve describes the running 
coupling constant in the physical scheme. For comparison, we 
plot in Fig.~2 this curve together with the 
curve for the running parameter for the Minimal Subtraction 
scheme, for large value of $t$, where a Landau pole shows up. 
It is not easy to visualize a difference between both 
cases, which look almost identical. Actually, the curve 
for the running parameter in the physical scheme is shifted 
a little bit to the right. This difference can be made clear 
if we increase the plot scale in the region around $t\sim 5900$, 
as illustrated in Fig.~3. Here, one can see 
actually two vertical lines (the dashed 
one for the MS scheme), indicating two different Landau 
poles. They are, however, very close, since the corresponding 
values for $t$ differ by less than $0.02\%$. The situation 
is very similar to the one described earlier for the 
scalar field theory \cite{bexi}.


\vskip 5mm
\begin{figure}
\includegraphics[width=0.5\textwidth]{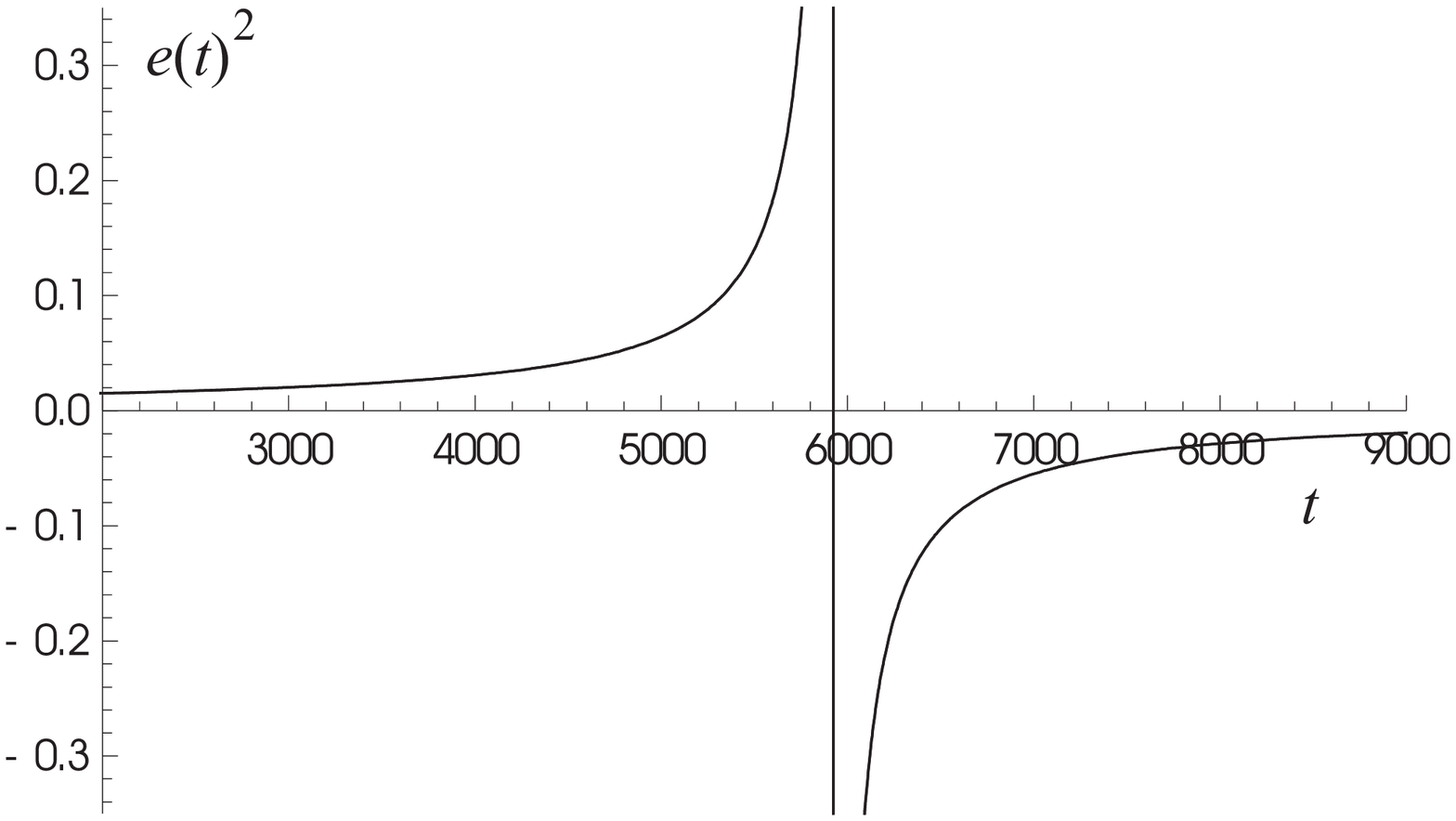}
\begin{quotation}
\caption{The curves for the running parameter $e(t)^2$ seem 
to coincide for both MS and physical scheme (in the last one 
we use the prescription corresponding to the form factor $k_1$). 
Actually, both curves are slightly different as well as the 
Landau pole (see Fig.~3). We have used $e(0)=0.1$.}
\end{quotation}
\label{Figura2N}
\end{figure}

\vskip 5mm
\begin{figure}
\includegraphics[width=0.5\textwidth]{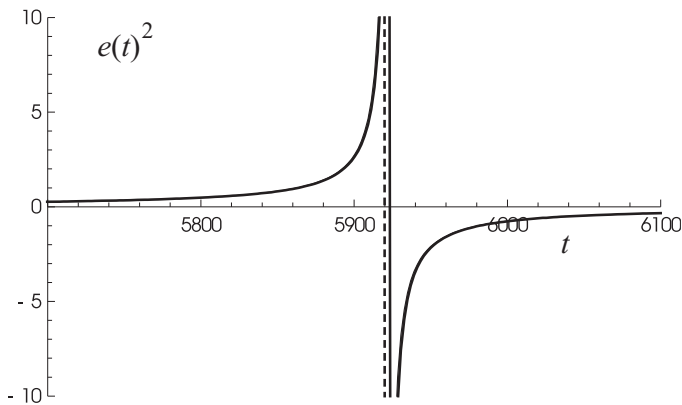}
\begin{quotation}
\caption{The same plot as in Fig.~2,
viewed with more detail around $t\sim 5900$. Here the two 
vertical lines indicate different Landau poles. The dashed 
line corresponds to the MS scheme. The value for $t$ 
corresponding to the poles differs from each other by less 
than $0.02\%$.}
\end{quotation}
\label{Figura3N}
\end{figure}

The running of the effective electromagnetic charge is 
shown in Fig.~4, where we plot $e^{-2}$ versus
$t = \ln (p/m)$. The plot for the $\overline{\rm MS}$ 
case is a straight line, as it has to be. The plot for the 
momentum subtraction scheme represents two straight lines 
in the asymptotic (IR and UV) limits with the smooth
transition between them in the intermediate region. It is 
easy to note that the curves are very similar to each other
in the UV region, that is for $t \gg 1$ (in fact this 
feature holds already for $t \geq 2$). At the same time, 
even in the UV the two plots do not coincide
and are represented by two parallel straight lines with 
slightly different initial points (at $t=0$, that means 
$p=\mu_0$ and $\mu=\mu_0$. The effect of the effective 
UV shift of initial point will take place not only in QED,
but also in QCD and EW sectors of the SM and beyond. 
This effect, despite being very small, may have interesting 
applications. For example, one can take into account the
effective shift of the initial values of the couplings
when calculating the UV effects in gauge theories via the 
renormalization group (see, e.g., \cite{QCD-UV-RG}). 
The calculations of this sort are quite relevant, 
including for the perspective LHC physics. Taking
the effect of the masses of the quantum fields into 
account may, in principle, improve the precision of 
the results. As a particular case, the effect of the initial 
data shift can be also observed in the supersymmetric models 
such as MSSM and its extensions. As a result, there may be a 
slight violation of the exact convergence of the running 
coupling constants $g$, $g^\prime$ and $g_s$ which will 
likely form a small triangle rather than meet in a single 
point. This feature of the massive theories has been already
discussed in \cite{Brodsky}, but the effect can be done 
more explicitly when using analytic expressions for 
the $\be$-functions. 

\vskip 5mm
\begin{figure}
\includegraphics[width=0.5\textwidth]{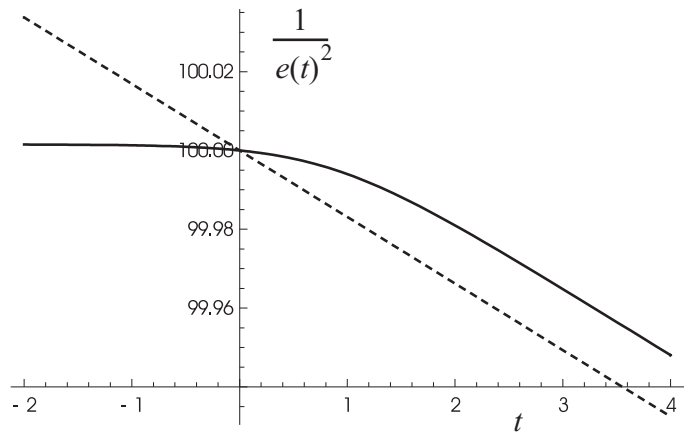}
\begin{quotation}
\caption{Curves for $e(t)^{-2}$. The dashed line corresponds 
to the MS scheme and the continuous line for the momentum
subtraction scheme. For higher $t$, both curves are 
parallel straight lines. Substantial difference between the 
two plots takes place for $t < 1$.}
\end{quotation}
\label{Figura3aN}
\end{figure}

\section{On the conformal violation in massive case}

In the previous sections we have considered the existing
interface between the quantum contributions coming from the
massive and massless field loops. In particular, we have 
seen that, in the high energy limit, the form factors and the 
$\be$-functions tend to the ones in the massless case, that 
means they become close to the ones in the minimal 
subtraction scheme of renormalization ($\overline{\rm MS}$, 
in our case). As we already learned in section 2, in this 
situation the violation of local conformal symmetry occurs
due to anomaly and the corresponding EA can be presented 
in a closed form (\ref{em-1-loop}).

On the contrary, the low-energy quantum effects are 
characterized by the phenomenon of decoupling \cite{AC}. 
As a result, in the IR limit the quantum effects are 
quadratically suppressed and one can rely on the classical 
Maxwell equations. 
An interesting question is what is the difference between the 
violation of local conformal symmetry in the ``far UV'' limit 
and in the ``far IR'' limit, where we have a remnant quantum 
effects. 
Let us emphasize that there is no contradiction in having 
two different forms for the violation of conformal symmetry, 
because they correspond to the two different physical 
situations.  

Let us first compare the dependences on the conformal 
factor in the two symmetry-violating terms. For the 
sake of simplicity we can take $\si=const$, that means
the global conformal symmetry. The situation for the 
local case will not be very different. In the 
UV limit, exactly as in the purely massless case we 
have, according to the Eq. (\ref{log box}), the linear 
dependence on $\si$. Of course, the same result follows 
from the complete expression (\ref{em-1-loop}) derived 
for the precisely-massless case. On the contrary, in the 
situation of IR decoupling, we can take the lowest order
terms directly from the $a_3$-coefficient. In 
this case one meets the $F\Box F$-type expression 
(\ref{decouple}) and, also, $RFF$-type terms. In both cases 
the $\,\si$-dependence is exponential, that means the 
symmetry is violated by the terms which have a scaling law 
$\,e^{-2\si}$. So, it looks like the scaling of the 
symmetry-violating terms is even stronger for the 
low-energy sector. However, this is nothing but a 
wrong impression. 

One can view the situation from a different position, if 
analyzing this question using the calculated expression
(\ref{FF1}). Then the physical sense of decoupling in the 
low-energy limit becomes much more explicit. If we compare 
the $F\Box F$-type expression (\ref{decouple}) and the 
classical term $F^2_{\mu\nu}$, it is clear that the 
former has an extra factor of $p^2/m^2$, where $p^2$
is the square of the momentum of the photon and the 
$m^2$ is the square of the mass of the electron. In 
the low-energy limit $p^2 \ll m^2$, hence we meet 
the simplest (and very clear) form of the Appelquist 
and Carazzone \cite{AC} quadratic decoupling law. In 
the case of $RFF$-type terms, for most of the physical 
situations, the decoupling is even much stronger. The
reason is that the scalar curvature is proportional to 
the square of the typical energy of the gravitons and 
this energy is much smaller than the one of the photon. 
For instance, in the cosmological setting we have 
$R \propto H^2$, where $H$ is the Hubble parameter. 
In the present-day Universe the corresponding values are
$H_0 \approx 10^{-42}\,GeV$ for the Hubble parameter
and $\ep = \sqrt{p^2} \approx 10^{-12}\,GeV$ for the 
energy of the CMB photon. Similar relation between the
two quantities holds during most of the evolution 
of the Universe. In this 
situation, the most important low-energy contribution 
to the violation of conformal symmetry comes from the 
(\ref{decouple}) term, which has a nontrivial flat 
limit.

\section{Discussions and summary}
  
The two distinct approaches to the derivation of one-loop
quantum corrections to the photon sector of the curved-space
QED have been explored. First, we derived the anomaly-induced 
action, coming from the integration of conformal anomaly. 
The effective action which we gain in
this way has an enormous advantage of being exact, in 
the sense it is not related to some particular form of 
series expansion, except the one into the loops (taken in
the first order). The representations obtained here can 
serve, in principle, as a consistent background for the 
investigation of quantum processes in the early universe, 
when the masses of quantum matter (fermionic) fields are
irrelevant. The anomaly-induced action has yet another 
advantage of being scheme independent, because it is 
based on the Minimal Subtraction renormalization scheme 
and is, after all, controlled by the one-loop divergences. 

On the other hand, in the massive case there is the 
violation of conformal symmetry coming from the form factor 
in the electromagnetic sector of QED in curved space-time.
This calculation is based on the physical  renormalization 
scheme and hence it is supposed to be more adequate in the 
later universe, when the masses of the fields play an 
important role. The price one has to pay for a more 
consistent physical approach is related to the restricted 
power of the available calculational methods, which are 
equivalent to the use of common Feynman diagrams for the 
linearized metric perturbations on the flat space-time 
background. The corresponding result which we obtain 
here includes terms which are quadratic in $F_{\mu\nu}$
and may also depend on the curvature tensor. It is 
complete in the case of a flat space-time background, 
but it is not supposed to be a complete one for the 
curved space-time, where we can expect many higher order
in curvature corrections which can not be calculated 
exactly. An interesting point is that we have found that 
the quantum correction depends on the choice of the 
calculational scheme \cite{multi}. Thus we have proven 
the existence of the nonlocal and renormalization 
independent MA in quantum field theory. One of the 
consequences of this anomaly is the ambiguity in the 
prediction of the Appelquist and Carazzone theorem 
\cite{AC}, which provides two different coefficients 
of the quadratic decoupling law at low energies. 

Since the quantum corrections in the electromagnetic 
sector include some ambiguity in the IR region, one 
should ask which one of the two schemes gives a correct 
result. In our opinion the advantage should be given 
to the one derived through the operator ${\hat H}^*_1$,
because it is more natural and preserves gauge invariance. 
However, it is worthwhile to be aware of the ambiguity 
which is a manifestation of a typical property 
for the off-shell effective action in quantum field
theory. In the present case this ambiguity becomes 
essential due to the presence of an external 
gravitational field. One can note that the simpler
form of quantum correction (\ref{em-1-loop}), derived 
via conformal anomaly, is also ambiguous due to the 
presence of an arbitrary functional $S_c$. Of course, 
the two kind of ambiguities are unrelated, but they 
can be seen as manifestations of a general 
feature of effective action.

\section*{Acknowledgments}
\noindent

I.Sh. is grateful to I. Buchbinder and B. Guberina for 
stimulating conversations on the subject of scheme 
dependence and to J. Sol\`a for the explanation 
concerning $\overline{\rm MS}$-based renormalization 
group in particle physics. 
Authors are grateful to CNPq, FAPEMIG and FAPES for support. 
The work of I.Sh. has been also supported by ICTP. 

\section*{Note Added.}

After this paper was resubmitted, we learned about the 
recent paper \cite{Mottola-08} (see also \cite{Cori})
where the same problem was 
considered using technically different approaches. We would 
like to cite this paper, especially because the parts of 
conformal anomaly-induced action are different only due 
to the choice of parametrization for the auxiliary fields. 
Also we would like to point out that, despite of the 
mentioned differences, the results are qualitatively
similar. In particular, in both approaches the conformal 
anomaly can be partially restored from the calculations of  
massive loop in the massless limit. 


\newpage

\end{document}